\documentstyle[11pt]{article}

\newcommand{\la}{\lambda}	
\newcommand{\bt}{\beta}
\newcommand{\beg}{\begin{equation}}
\newcommand{\enq}{\end{equation}}
\newcommand{\sig}{\sigma}

\newcommand{\ap}{\alpha}

\newcommand{\ep}{\varepsilon}

\newcommand{\Lam}{\Lambda}
\newcommand{\sfrac}[2]{\sqrt{\frac{#1}{#2}}}
\newcommand{\e}{\varepsilon}
\newcommand{\atanh}{\tanh^{-1}}
\newcommand{\half}{\frac{1}{2}}
\newcommand{\cost}{{\rm constant}}
\newcommand{\hy}{{\;_2F_1}}
\newcommand{\A}{{\rm A}}
\newcommand{\B}{{\rm B}}
\newcommand{\C}{{\rm C}}
\newcommand{\eff}{{\rm eff}}

\renewcommand{\thesection}{\arabic{section}.$\!\!\!\!\!$}

\begin{document}

\title{Some Eigenvalue Distribution Functions 
of the Laguerre Ensemble}
\author{Y Chen$^{\dag,1}$
and S M Manning$^{\S,2}$\\
        $^{\dag}$Department of Mathematics, Imperial College\\
        180 Queen's Gate, London SW7 2BZ\\
        $^{\S}$Department of Theoretical Physics, Oxford University\\
        1 Keble Road, Oxford OX1 3NP}
\date{\today}
\maketitle

\begin{abstract}
In this paper we revisit the smallest-eigenvalue distribution of the
Laguerre ensemble by presenting in closed form certain integrals
obtained previously.  With this information we compute, using Dyson's
continuum approximation, the two-smallest-eigenvalue distribution of
the Laguerre ensemble.  Higher-order contributions to the free
energies describing these two probabilities in certain scaling limits
for $\bt\neq 2,$ which can be interpreted in this context as entropic
effect, are found.
\end{abstract}
e-mail:$^1$y.chen@ic.ac.uk, $^2$s.manning1@physics.oxford.ac.uk
\newpage

\section{Introduction}

In a previous paper \cite{Chen1} the smallest-eigenvalue distribution
function for the Laguerre Ensemble was found via Dyson's \cite{Dyson}
continuum approximation in a certain (double-) scaling limit (to be
described below). The continuum approximation treats the eigenvalues
of an ensemble of $N\times N$ random matrices by a continuous fluid in
the limit large $N$, with its thermodynamic free energy completely
characterized as a functional of the fluid density.  Although a
rigorous proof of the general validity of such an approximation is not
available, we have confidence in the robustness of this approach,
since results obtained using it \cite{Dyson,Chen1,Chen2,Chen3,Chen4}
compare favourably with those obtained from exact analyses, whenever
such analyses are available.

Quite generally, the probability functions which concern us can be
computed as the change in the free energy when the support of the
fluid is perturbed.  In most situations the determination of the
smallest-eigenvalue distribution and other allied quantities can be
posed as the following problem in equilibrium statistical mechanics:
what is probability, ${\rm prob}(R)$, of finding a bubble of radius
$R$ due to thermal fluctuations in the bulk of a fluid at temperature
$T$? This important quantity measures the degree of correlation in the
fluid.  As the creation of a void requires energy, such events are
relatively rare.  Of particular interest is the behaviour of this
probability when the bubble is large. A physical argument that gives a
feel for the answer goes as follows: creating a bubble of radius $R$
in $d-$Euclidean dimensions requires a volume energy $\propto R^{d}$
and a surface energy $\propto R^{d-1}$.  An application of the
Boltzmann principle then suggests that 
\begin{equation}
{\rm prob}(R)\sim {\rm e}^{-c_1\;R^{d}-c_2\;R^{d-1}},
\end{equation}
where $c_1$ and $c_2$, being respectively the volume and surface
energies divided by $k_{\rm B}T$, are positive constants independent
of $R$.  Eq.(1) is valid for a bubble with a radius much larger than
the coherence length.  Thus, ${\rm prob}(R)\sim {\rm e}^{-{\cost\times
R}}$ in one dimension. For a fluid with short-ranged interactions
Eq.(1) is expected to hold, but it should fail for a fluid with
long-ranged interactions as is the case for a charged system, because
the volume and surface energies are then no longer well defined.  That
such a consideration is pertinent to random-matrix problems is due to
the appropriate measure in the space of $N\times N$ matrices
consistent with symmetries. The invariant measure, according to a
classification theorem of Dyson \cite{Mehta} is
$$P(M)dM={\rm exp}\left[-{\rm tr}u(M)\right]dM
=\prod_{1\leq a<b\leq N}|x_a-x_b|^{\bt}\prod_{c=1}^{N}
{\rm exp}\left[-u(x_c)\right]
dx_c\;dG,
$$
where $x_a$ are the eigenvalues and $dG$ is the Haar measure of the
symmetry group that diagonalizes the matrices with orthogonal, unitary
and symplectic symmetries, corresponding to $\bt=1,2,4,$ respectively.

The probability that an interval $J$ (a subset of {\bf R})
is free of eigenvalues, also known as the Gap Formation Probability
(GFP), is then
\begin{equation}
E_{\bt}[J]=\frac{\prod_{c=1}^{N}\left(\int_{\bar J}d\mu(x_c)\right)\;
{\rm e}^{-W(x_1,\ldots,x_N)}}{\prod_{c=1}^{N}
\left(\int_{{\bar J}\cup J}d\mu(x_c)\right)
{\rm e}^{-W(x_1,\ldots,x_N)}},
\end{equation}
where ${\bar J}$ is the compliment of $J$, ${\bar J}\cup J$ is 
natural support of the eigenvalues, $d\mu(x):={\rm exp}[-u(x)]dx$ and
\beg
W(x_1,\ldots,x_N):=-\bt\sum_{1\leq a<b\leq N}\ln|x_a-x_b|.
\enq
Insight into this distribution can be gained by considering the $x$'s
as the spatial coordinates of the particles and $u(x)$ can then be
interpreted as a form of ``confining'' potential holding together the
logarithmically-repelling charged particles (all carrying charge $+1$)
moving in one-dimension.  Under the combined influence of the
confining and repelling forces the charges will attempt to settle down
to an optimal configuration with small fluctuations around it.  In the
limit large $N$ we can treat this system as a continuous fluid using
techniques from electrostatics and thermodynamics.  From the above
consideration, the GFP can be written as
\beg 
E_{\bt}[J]={\rm e}^{-\delta F[J]},
\enq 
where
$$
\delta F[J]:=F[{\bar J}]-F[{\bar J\cup J}],
$$
is the free energy of $N$ charges confined to region ${\bar J}$ minus
the free energy of $N$ charges in the natural support of $d\mu.$ As
indicated above, the heuristic formula---Eq.(1)---for the GFP fails
for systems with long-ranged interaction.  Instead, an alternative
argument peculiar to charged systems which provides a qualitatively
correct estimate for a large interval can be formulated as follows.
An appropriate counter charge distribution is introduced externally to
neutralize the region from which we wish to have the eigenvalues
removed.  If $F_Q$ is the free energy of the system with the addition
of the counter charge ($Q$) and $F_{Q=0}$ the free energy of the
undisturbed system then, according electrostatics, $F_Q$ differs from
$F_{Q=0}$ by an amount equal to the charging energy, {\it i.e.}
\beg 
F_Q\sim F_0+F_{\rm charging},
\enq
where
\begin{eqnarray}
F_{\rm charging} & = & \frac{Q^2}{2C}
\propto {\cal N}^2[J],\;\;\;C={\rm capacitance}
\nonumber \\
& = &\left[{\rm the\;total\;number\;of\;
eigenvalues\;excluded\;from\;}J\right]^2=
\left[\int_J\;dx\;\sig(x)\right]^2,
\end{eqnarray}
and
$\sig(x)$ is the eigenvalue/charge density
\footnote{Exact Painlev\'e analysis \cite{Tracy} for large $|J|$ and
for $\bt=2,$ whenever such treatments are available, show that there
is a sub-leading term $\sim \ln{\cal N}[J].$ This is not obtainable
from the charging energy arguments. However, as was shown previously
\cite{Chen1} and will be shown later for the Laguerre Ensemble, the
logarithmic correction is obtained in the continuum approximation.
This is due to the fact that the density of the Laguerre Ensemble is
very large at the ``hard edge.''}.  Numerous examples supporting
Eq.(6) can be found in \cite{Chen2,Chen3}.  A further example that
deals with the Laguerre ensemble is given in Appendix A.  We mention
here that a screening theory of the charged fluid gives a physical
justification \cite{Chen2} of Eq.(2), although we do not know of a
proof of the general validity of this relationship.

In this paper we will use Dyson's continuum approximation, so let us
first briefly outline what that entails.  Dyson showed that the
functional dependence of the free energy on the charge density is
given by the functional
\begin{equation}
F[\sigma]:=\int_K\;dx\;u(x)\sigma(x)-\frac{\bt}{2}
\int_Kdx\int_Kdy\;\sigma(x)\ln|x-y|\sigma(y)+
\left(1-\frac{\bt}{2}\right)\int_K\;dx\sigma(x)\ln[\sigma(x)],
\label{eq:continuum}
\end{equation}
where the fluid is confined to the region $K$ and the third term on
the right-hand side of Eq.(\ref{eq:continuum}) came out of Dyson's
careful analysis and will be referred to as the entropic term.  Given
this functional, then the equilibrium density $\sigma(x)$ is
determined by minimization of $F[\sigma]$ with respect to the $\sigma(x)$
subject to the normalization constraint
\begin{equation}
\int_{K}\;\sigma(x)dx = N, 
\end{equation}
$N$ being the number of charges, which is equal to the rank of the
underlying matrices.  Using Lagrange's method to handle this
constrained minimization, which is equivalent to introducing a
chemical potential, $A$, shows that the integral equation satisfied by
the ground-state density is
\begin{equation}
u(x) - \beta\int_{K}dy\;\;\sigma(y)\ln|x-y| + 
	\left(1 - \frac{\beta}{2}\right)\ln\;\sigma(x) = A.
\label{eq:basic_eqn}
\enq 
Therefore the free energy at equilibrium and with precisely $N$ charges
contained in $K$ is 
\begin{equation}
F[K] =\frac{1}{2}NA+\frac{1}{2}\int_K\;dx\;u(x)\;\sigma(x)
-\left(1-\frac{\beta}{2}\right)\int_K\;dx\;\sigma(x)\ln\sigma(x),
\end{equation}
the calculation of which for various different cases will form the
core of this work.

This paper is organized as follows: in Section 2 we revisit the
Laguerre ensemble and compute the probability that the interval
$[0,a]$ is free of eigenvalues by utilizing closed form expressions of
certain integrals that were previously determined approximately and
supply in a (double-) scaling limit the change in free energy
including the entropic $(\beta\neq 2)$ contribution; with this
information we then present in Section 3 the
two-smallest-eigenvalue-distribution function for the Laguerre
ensemble; in section 4 we present our conclusions and discuss the
extension of our method to the $m$-eigenvalue problem; details of the
calculation of certain key integrals are given in the appendicies,
along with a note on the charging energy of the Laguerre ensemble.

\section{The Laguerre Potential Revisited}
\label{sec:Lag_rev}

The confining potential $u(x)$ for the Laguerre Ensemble is
\begin{equation}
u(x)=x-\alpha\ln x,\;\;x\geq 0{\;\;\rm and\;\;}\alpha>-1.
\end{equation}
The change in free energy is computed as
\beg
\delta F=F[b,a]-F[b,0],
\enq
where $F[b,a]$ is the free energy where all $N$ charges are supported
on $a<x<b.$ This would require the determination of the fluid density
$\sig(x)$ which satisfies the singular integral equation\footnote{This
equation is found by differentiating Eq.(9) with respect to $x,$ for
$\bt=2.$ The $\bt\neq 2$ case will be considered later.}
\begin{equation}
\frac{du}{dx}=\bt\;{\rm P}\int_{a}^{b}dy\frac{\sig(y)}{x-y},
\;\;a<x<b,
\end{equation}
the general solution of which for the ground state is
\footnote{$c/{\sqrt {(b-x)(x-a)}}$ ($c$ a constant) solves 
the homogeneous part of Eq.(\ref{eq:gen_soln}), but including this
would clearly increase the free energy and so it must be excluded from
the solution for the ground state density.}  
\begin{equation}
\sig(x)=\frac{1}{\pi^2\bt}{\sqrt {\frac{b-x}{x-a}}}
{\rm P}\int_{a}^{b}\frac{dy}{y-x}{\sqrt {\frac{y-a}{b-y}}}\;
\frac{du(y)}{dy},\;\;a<x<b.
\label{eq:gen_soln}
\end{equation}
For the Laguerre potential the fluid density is therefore
\begin{equation}
\sigma(x) = =\frac{1}{\pi\beta}\sfrac{b-x}{x-a}
\left[1-\frac{\alpha}{x}{\sqrt {\frac{a}{b}}}\right],
\label{eq:sigma_Lag}
\end{equation}
where we require that $\sqrt{ab}>\alpha$ so as to ensure that
$\sigma(x)$ is positive semi-definite.  The normalization condition
now reads
\beg
N=\frac{b-a}{2\bt}+\frac{\alpha}{\bt}\left( \sfrac{a}{b}-1 \right),
\label{eq:N_Lag}
\enq
which gives $b$ as a function of $N$ and $a$ via a solution of the 
following cubic equation (found from a slight rearrangement of
Eq.(\ref{eq:N_Lag})):
\begin{equation}
b^3-2(\nu+a)b^2+(\nu+a)b-4\alpha^2=0,\;\;\nu:=2\bt N+2\alpha.
\end{equation}
For sufficiently large $N$ this has a positive discriminant and
therefore three real solutions, determinable by the trigonometric
method \cite{B_and_M}.  Since it is a simple matter to convince oneself
that $b$ decreases on increasing $a$ and that the leading order
contribution in $a$ should therefore be negative it follows that the
root which we require is
\begin{equation}
b=\frac{2(\nu+a)}{3}\left[1+\cos\left(\frac{1}{3}\cos^{-1}
\left[-1+\frac{54\alpha^2a}{(\nu+\bt)^3}\right]-\frac{2\pi}{3}\right)
\right].
\end{equation}

\subsection{The Free Energy}

Given $b$ and $\sigma(x)$ we can now determine the free energy which
is given by the sum of the chemical potential contribution 
$F_{\rm chem} := NA/2 = \left(\frac{\nu-\alpha}{4\bt}\right)A$ and
that due to interaction, {\it i.e.},
\begin{equation}
F_{\rm int}:=\frac{1}{2}\int_{a}^{b}
\sig(x)u(x)dx.
\end{equation}
To compute the chemical potential we first drop the $\beta \neq 2$
correction in Eq.(\ref{eq:basic_eqn}) (we will return to this term in
Section \ref{ssec:beta_neq_2}) and then evaluate it with $x
\rightarrow b$ giving
\begin{equation}
A = b-\alpha\ln b-\frac{b-a}{2}\ln\left(\frac{b-a}{4e}\right)
+\alpha\left[1-{\sqrt {\frac{a}{a}}}\right]+ \alpha{\sqrt
{\frac{a}{b}}}\;I_1\left(\frac{a}{b-a}\right),
\end{equation}
where $I_{1}(x)$ is tabulated along with the other integrals
$I_{j}(x)$, $j=2\ldots 4$ in the Table, whilst the details of their
computations are confined to \ref{app:integrals}  Using the Table we
have
\begin{equation} 
F_{\rm chem}=
\frac{\nu-\alpha}{4\bt}\left[a-\left(\frac{b-a}{2}\right)
\ln\left(\frac{b-a}{4e}\right)-\ap{\sqrt {\frac{a}{b}}}\ln
\left(\frac{b-a}{4}\right)-2\ap\tanh^{-1}{\sqrt
{\frac{a}{b}}}\;\right].  
\end{equation}
The interaction contribution to the free energy is slightly more
complicated;
$$
F_{\rm int}=\frac{(b-a)^2}{16\bt}+\frac{b-a}{4\bt}
\left[a-\ap{\sqrt {a}{b}}-\ap\ln(b-a)\right]-\frac{\ap(b-a)}{2\bt}
I_3\left(\frac{a}{b-a}\right)$$
\beg
+\frac{\ap^2\ln(b-a)}{2\bt}
\left(1-{\sqrt {\frac{a}{b}}}\;\right)+\frac{\ap^2}{2\bt}
I_2\left(\frac{a}{b-a}\right),\enq
where $I_2(x)$ and $I_3(x)$ are in the Table.

\subsection{The Scaling Limit of $F[b,a]$}

From the expressions for $F_{\rm chem}$ and $F_{\rm int}$ we shall now
obtain the free energy in the double scaling limit, {\it i.e.}, the
limit where $N$ (and therefore $\nu$) $\rightarrow\infty$ and
simultaneously $a\rightarrow 0,$ such that the combination $s:=\nu a$
is finite. Under the above limit the free energy will become a
function of $N$ plus a function of $s$ only; other terms of ${\rm
O}(a^{\mu}/\nu^{\delta}),\;\;\mu,\delta>0$ are discarded. Thus,
expanding the free energies in $a,$ we get
\begin{equation}
F[b,a]\sim F_0(N)+\frac{\ap^2}{2\bt}\;\ln s+\frac{2\ap}{\bt}\;
{\sqrt s}+\frac{s}{2\bt},\;\;F_0(N)\sim -\frac{\bt}{2}N^2\ln N.
\label{eq:F_gap_no_ent}
\end{equation}
The GFP is thus given by
\beg
E_{\bt}(s)\sim 
\frac{{\rm exp}\left[-\frac{s}{2\bt}+\frac{2\ap}{\bt}{\sqrt s}
\right]}
{s^{\frac{\ap^2}{2\bt}}},
\enq
in agreement with previously obtained results \cite{Chen1,Tracy}.

\subsection{The Entropic Contribution to the Free Energy for
$\beta\neq 2$}
\label{ssec:beta_neq_2}

The results we have so far obtained are strictly only true for $\beta
= 2$, as for $\bt\neq 2$ there is an extra entropic contribution to
the free energy and in the previous sections we have ignored it.
However, if we assume that the factor $(1 - \beta/2)$ is small, then
to lowest order in this factor we can obtain the total free energy by
adding in the term
\begin{equation}
F_{\rm ent}:=\left(1-\frac{\bt}{2}\right)\int_{a}^{b}\;\sig(x)
\;\ln\sig(x)\;dx 
=\left(1-\frac{\bt}{2}\right)f_{\rm ent}
\label{eq:Fent}
\end{equation}
to $F$, the density being as per Eq.(\ref{eq:sigma_Lag}), and, with
$\Lambda:=\frac{\alpha}{b-a}\sfrac{a}{b}$ and $x:=\frac{a}{b-a}$,
\begin{equation}
f_{\rm ent}:=-\ln(\pi\bt)N+f_{\rm ent}^{(1)}+f_{\rm ent}^{(2)},
\label{eq:f_split}
\end{equation}
where
\beg
f_{\rm ent}^{(1)}:=
\frac{b-a}{\pi\bt}
\int_{0}^{1}dt{\sqrt {\frac{1-t}{t}}}\left(1-\frac{\Lambda}{t+x}
\right)\ln{\sqrt {\frac{1-t}{t}}},\enq
and
\beg
f_{\rm ent}^{(2)}:=\frac{b-a}{\pi\bt}\int_{0}^{1}dt
{\sqrt {\frac{1-t}{t}}}\left(1-\frac{\Lambda}{t+x}\right)\ln
\left(1-\frac{\Lambda}{t+x}\right).
\enq
We start with $f_{\rm ent}^{(1)}$ which can be straightforwardly
reduced to
\begin{eqnarray}
f_{\rm ent}^{(1)} & = & \frac{b-a}{\bt}-\frac{\Lam(b-a)}{2\bt}
	\left[I_1(x)-I_4(x)\right]	\\
	& = & \frac{b-a}{\bt} + 
		\frac{\ap}{2\bt}\ln\left(\frac{a}{b}\right).
\label{eq:f_ent_2}
\end{eqnarray}
Thus, in the scaling limit, we find that
\begin{equation}
f_{\rm ent}^{(1)}\sim \frac{\nu}{\bt}+\frac{\ap}{2\bt}\ln s+
{\rm O}\left(\frac{s^{1/2}}{\nu}\right).
\label{eq:f_gap_ent_1}
\end{equation}
Concerning $f_{\rm ent}^{(2)}$, we first observe that $\Lambda/x < 1$,
because of the constraint $\alpha < \sqrt{ab}$ imposed to ensure that
the density is non-negative, which also makes certain that logarithm
in $F_{\rm ent}$ never leads to imaginary terms.  Thus, for $t \geq
0$, $\frac{\Lambda}{t + x} < 1$ and we can therefore evaluate
$f^{(2)}_{\rm ent}$ by expanding the logarithm in
Eq.(\ref{eq:f_ent_2}) as a power series in $\frac{\Lambda}{t+x}$.  In
the scaling limit we find that 
\beg f_{\rm ent}^{(2)}\sim
-\ap+\frac{1}{\bt}
\sum_{n=1}^{\infty}\frac{\ap^{n+1}}{4^n\left[(n+1)!\right]^2}\;
\left(1+\frac{1}{n}\right)\;\frac{1}{s^{\frac{n}{2}}},
\label{eq:f_gap_ent_2}
\enq 
the details of the derivation being in \ref{app:entropic_term} 
We have, upon collecting Eqs.(\ref{eq:F_gap_no_ent}), 
(\ref{eq:f_gap_ent_1}) and (\ref{eq:f_gap_ent_2}),
\begin{equation}
E_{\bt}(s)\sim c\frac{{\rm exp}
\left[-\frac{s}{2\bt}+\frac{2\ap}{\bt}{\sqrt s}\right]}
{s^{\frac{\ap^2}{2\bt}+\frac{\ap(2-\bt)}{4\bt}}}
\left[1+{\rm O}\left(\frac{1}{\sqrt s}\right)\right]
\label{eq:GFP},
\end{equation}
$c$ being a constant.

To conclude this section we note that from a probability argument we
can also obtain the smallest-eigenvalue distribution
$p_{\beta}^{(1)}(s)$, as it is related to $E_{\beta}(s)$ by
\cite{gap_prob} 
\begin{equation}
p_{\bt}^{(1)}(s)=-\frac{dE_{\bt}(s)}{ds}.
\label{eq:p_rel_E}
\end{equation}
Thus, from Eqs.(\ref{eq:GFP}) and (\ref{eq:p_rel_E}), we find that
\begin{equation}
p_{\beta}^{(1)} \sim \frac{c}{2\beta}\left(
	1 - \frac{2\alpha}{s^{\half}} + \left[
		\alpha + 1 - \frac{\beta}{2}\right]\frac{\alpha}{s}
	\right)\frac{\exp\left[-\frac{s}{2\beta} + 
			\frac{2\alpha}{\beta}s^{\half}\right]}
		{s^{\frac{\alpha^{2}}{2\beta} + 
		\frac{\alpha(2-\beta)}{4\beta}}}\left[1 + 
			O\left(s^{-\half}\right)\right].
\end{equation}
\section{The Two-Smallest-Eigenvalue Distribution}

In this section we will calculate, within the continuum approximation,
$p^{(2)}_{\beta}(x_{1},x_{2})$, the distribution function of the two
smallest eigenvalues $x_{1}$ and $x_{2}$ ($x_{2} \geq x_{1}$), given
that the eigenvalues are distributed according to the joint
probability distribution function
\begin{equation}
p(x_{1},\ldots,x_{N}) = \frac{1}{Z}\prod_{a=1}^{N}
e^{-u(x_{a})}\prod_{1 \leq b < c \leq N}|x_{b}-x_{c}|^{\beta},
\label{eq:eval_dbn}
\end{equation}
where $Z$ is the partition function.  To calculate
$p^{(2)}_{\beta}(x_{1},x_{2})$ we fix $x_{1}$ and $x_{2}$ and
integrate $p(x_1,\ldots, x_N),$ with respect to $x_3,\ldots , x_N$
from $x_{2}$ to $\infty$, then
\begin{eqnarray}
\lefteqn{p_{\beta}^{(2)}(x_{1},x_{2})
=\frac{N(N-1)}{2Z}|x_{1}-x_{2}|^{\beta}e^{-u(x_{1})-u(x_{2})}
\times}&& \nonumber \\
&&\prod_{a=3}^{N}\left[\int_{x_{2}}^{\infty}dx_a\;e^{-u(x_a)}
|x_{1}-x_{a}|^{\bt}|x_{2}-x_{a}|^{\bt}\right]
\prod_{3\leq b<c\leq N}|x_b-x_c|^{\bt}.
\end{eqnarray}
Shifting the integration variables using $x_a=t_a+x_{2}$ and a
relabelling the indices, we get
\begin{eqnarray}
\lefteqn{p_{\bt}^{(2)}(x_{1},x_{2}) = {\cal C}
	|x_{1}-x_{2}|^{\bt}e^{-u(x_{1})-u(x_{2})}
	\times} && \nonumber \\
&& \prod_{a=1}^{N-2}\left[\int_{0}^{\infty}dt_a\;e^{-u(t_a+x_{2})}\;
|x_{1}-x_{2}-t_a|^{\bt}|t_a|^{\bt}\right]\prod_{1\leq b<c\leq N-2}
|t_b-t_c|^{\bt},
\label{eq:p^2}
\end{eqnarray}
where ${\cal C} = \frac{N(N-1)}{2Z}$.  The multiple integral after the
trivial factors will be evaluated using the continuum approximation
described above.  This is particularly interesting since the
logarithmic (one-particle) factors already appear here.  We expect,
from the results obtained in Section 2 that there will be an analogous
logarithmic correction to the free energy. It is convenient to next
transform to the centre-of-mass and difference coordinates, $\e =
x_{2}-x_{1} (\geq 0)$ and $R = (x_{1} + x_{2})/2$ respectively, and
then take $\bar{p}^{2}_{\beta}(\e,R) = p^{(2)}_{\beta}(x_{1},x_{2})$.
Then
\begin{eqnarray}
\lefteqn{\bar{p}_{\beta}^{(2)}(\e,R) = {\cal C}\e^{\bt}\;
{\rm exp}\left[
	-u\left(R-\frac{\e}{2}\right)-u\left(R+\frac{\e}{2}\right)
\right]\times} && \nonumber \\
&&
\int_{0}^{\infty}\left\{\prod_{a=1}^{N-2}dt_a\right\}
	\left[{\rm e}^{-u(t_a + R + \e/2)}\;
(t_a(t_a+\e))^{\bt}\;
\prod_{1\leq b < c \leq N-2}|t_b-t_c|^{\bt}\right].  
\label{eq:barp^2}
\end{eqnarray}
The multiple integral in Eq.(\ref{eq:barp^2}) can be
interpreted as the partition function for a charged fluid
of $N-2$ of particles in the effective external potential
$u_{{\rm eff}}(t;\e,R)$ defined by
\begin{equation}
u_{{\rm eff}}(t;\e,R) = u(t+R+\ep/2)-\bt\ln[t(t+\ep)].
\label{eq:u_eff}
\end{equation}
For reasons which we will explain later, the potential
with $\alpha = 0$ is much simpler than the $\alpha \neq
0$ case, so we will first calculate the asymptotic
expansion of $p^{(2)}_{\beta}$ for $\alpha = 0$.  We
therefore have
\begin{eqnarray}
\lefteqn{
	\bar{p}_{\beta}^{(2)}(\e,R) =
	{\cal C}\e^{\beta}{\rm e}^{-2R}
	\times} && \nonumber \\
&&
\int_{0}^{\infty}\left\{\prod_{a=1}^{N-2}dx_a\right\}
\left[\;{\rm e}^{-\left(x_{a} + R + \frac{\e}{2}\right)}
\left[x_a(x_a+\ep)^{\bt}\right]
\prod_{1\leq b<c\leq N-2}|x_b-x_c|^{\bt}\right].
\end{eqnarray}
The analogous integral equation for electrostatic equilibrium is,
\begin{equation}
A = x + R + \frac{\e}{2} 
	- \bt\ln x-\bt \ln(x+\ep)-\bt\int_{0}^{b}dy\sig(y)\ln|x-y|,
\label{eq:A_2eval}
\end{equation}
where $A$ is the chemical potential, or
\begin{equation}
1-\bt\left(\frac{1}{x}+\frac{1}{x+\ep}\right)=\bt\;{\rm P}\int_{0}^{b}
dy\frac{\sig(y)}{y-x},
\label{eq:int_eqn_2eval}
\end{equation}
following the method outlined in Section 2.  The
solution of Eq.(\ref{eq:int_eqn_2eval}) is
\begin{equation}
\sig(x)=\frac{1}{\pi\bt}{\sqrt {\frac{b-x}{x}}}
\left[1-\frac{\bt}{x+\ep}{\sqrt
{\frac{\ep}{b+\ep}}}\;\right], \;\;0<x<b.
\end{equation}
With the normalization $ \int_{0}^{b}dx\sig(x)=N-2,$ we find
\begin{equation}
N-1=\frac{b}{2\bt}+{\sqrt \frac{\ep}{b+\ep}},
\end{equation}
which is equivalent to the cubic equation
\footnote{Recognising that we could obtain an {\it
exact} solution for $b$ prompted us to look for
closed-form expressions for the integrals and hence to
re-examine the level-spacing distribution, the results
for which we presented earlier.}
\begin{equation}
X^3 - 2(\nu+\ep)X^2 + (\nu+\ep)^{2}X
	- 4\beta^{2}\e = 0,\;\;X:=b+\e,\;\;
\nu:=2\beta(N-1).
\label{eq:cubic_2eval}
\end{equation}
Again the discriminant of Eq.(\ref{eq:cubic_2eval}) is
positive for sufficiently large $N$ and so we have three
real roots.  The appropriate solution is,
\begin{equation}
b=\frac{2(\nu+\ep)}{3}\left[1+
\cos\left(\frac{1}{3}\cos^{-1}
\left(-1+54\bt^2\ep/(\nu+\ep)^3\right)-\frac{2\pi}{3}
\right)\right]-\ep.
\end{equation}
The analogous free energy is
\begin{equation}
F = \frac{(\nu-2\beta)A}{4\beta} + F_{{\rm int}},
\label{eq:F_2eval}
\end{equation}
where 
\begin{equation}
F_{{\rm int}} := 
	\frac{1}{2}\int_{0}^{b}dx\sig(x)u_{{\rm eff}}(x;\e,R),
\end{equation}
is the appropriate interaction energy.

\subsection{The Chemical Potential $A$}

To compute the chemical potential we take $x = b$ in
Eq.(\ref{eq:A_2eval}), then
\begin{equation}
A = \frac{\ln(4e)}{2}b + R + \frac{\e}{2} - 
\beta\ln[b(b+\e)] - \frac{b\ln b}{2} + \beta\ln b
\left[1 - \sfrac{x}{1 + x}\right] + \beta\sfrac{x}{1 + x}I_{1}(x),
\end{equation}
defining $x := \frac{\e}{b}$.  In the scaling limits $\nu\rightarrow
\infty$ and $\ep, R \rightarrow 0$ we find two suitable scaled
variables: $S_{1} := \nu\e$ and $S_{2} := \nu(R + \e/2)$ (for reasons
which will be made clear below), yielding
\begin{equation}
F_{{\rm chem}} \sim
-\frac{\nu^2}{8\beta}\ln\left(\frac{\nu}{4e}\right) - 
\frac{\nu}{4}\ln(4e)+\frac{\beta}{2}\ln\nu - 
\frac{{\sqrt S_{1}}}{2} + \frac{S_{2}}{4\beta} + 
{\rm O}\left(\frac{S_{1}^{1/2}}{\nu}\right)
\label{eq:Fchem_2eval}
\end{equation}
as the asymptotic expansion for $F_{{\rm chem}}$.
\noindent

\subsection{The Interaction Energy}

In order to simplify the notation we first split up $F_{int}$ into the
sum of two terms, $F_{1}$ and $F_{\ln}$, where
\begin{equation}
F_{1} := \frac{1}{2\pi\beta}
	\int_{0}^{1}dt{\sqrt {\frac{1-t}{t}}}
\left(b-\sfrac{\e}{b + \e}\frac{\beta}{t+x}\right)
	\left[bt + R + \frac{\e}{2} - 2\beta\ln b\right]
\end{equation}
and
\begin{equation}
F_{\ln} := -\frac{1}{2\pi}\int_{0}^{1}dt{\sqrt {\frac{1-t}{t}}}
\left(b - \sfrac{\e}{b + \e}\frac{\beta}{t+x}\right)
	\;\ln[t(t+x)].
\end{equation}
Evaluating these integrals and expanding the results we find that:
\begin{eqnarray}
F_{1} & = & \frac{b^2}{16\beta} - 
	\frac{b}{4}{\sqrt {\frac{\ep}{b+\ep}}}-\frac{\ep}{2}
	\left({\sqrt {\frac{\ep}{b+\ep}}}-1\right) + 
	\frac{N-2}{2}\left(R + \frac{\e}{2} - 2\beta\ln b\right)
\nonumber \\
	& \sim & \frac{\nu^2}{16\beta} - \frac{\nu\ln\nu}{2} -
	\frac{\sqrt S_{1}}{2} + \frac{S_{2}}{4\beta} + 
	{\rm O}\left(\frac{S_{1}}{\nu}\right)
\end{eqnarray}
and
\begin{eqnarray}
F_{\ln} &=&	\ln(2) b - 
\frac{b}{2}I_3(x)-\beta\ln b\left({\sqrt {\frac{x}{x+1}}}-1\right)
+\frac{\beta}{2}I_2(x)+\frac{\beta}{2}{\sqrt
{\frac{x}{x+1}}}I_4(x)
\nonumber \\
& \sim & \half\left(\beta\ln 2 - \half\ln(4e)\right)\nu - 2\beta\ln\nu +
\beta\ln S_{1} - {\sqrt S_{1}} +
{\rm O}\left(\frac{S_{1}^{1/2}}{\nu}\right).
\end{eqnarray}
The asymptotic expansion of $F_{\rm int}$ is thus
\begin{equation}
F_{\rm int} \sim F^{0}_{\rm int}(\nu) + \beta\ln S_{1} -
\frac{3}{2}\sqrt{S_{1}} + \frac{S_{2}}{4\beta}.
\label{eq:Fint_2eval}
\end{equation}

From Eqs.(\ref{eq:F_2eval}), (\ref{eq:Fchem_2eval}) and
(\ref{eq:Fint_2eval}) we therefore find that asymptotically
\begin{equation}
F(\nu;S_{1},S_{2}) \sim F^{0}(\nu) + \beta\ln(S_{1}) - 2\sqrt{S_{1}} +
\frac{S_{2}}{2\beta}.
\label{eq:F_aeq0_2eval}
\end{equation}
\subsection{The Entropic Contribution to the Free Energy}

As explained in section \ref{ssec:beta_neq_2} when $\beta \neq 2$
there are additional terms in the free energy, the lowest order
correction resembling an entropic contribution and is given by
Eq.(\ref{eq:Fent}).  We can once again split up $F_{\rm ent}$ as per
Eq.(\ref{eq:f_split}), with $f^{(1)}_{\rm ent}$ and 
$f^{(2)}_{\rm ent}$ as
before, except for that the prefactor is changed according to 
$\frac{b-a}{\pi\beta} \rightarrow \frac{b}{\pi\beta}$ and we now take
$\Lambda = \frac{\beta}{b}\sfrac{x}{1 + x}$.  It follows that we now
have
\begin{equation}
f^{(1)}_{\rm ent} \sim \frac{\nu}{\beta} - \ln\nu + \frac{1}{2}\ln S_{1}
	+ O\left(\frac{S_{1}^{\half}}{\nu}\right)
\end{equation}
and
\begin{equation}
f^{(2)}_{\rm ent} \sim -1 + 
	\sum_{n = 1}^{\infty}\frac{(2n)!}{[(n + 1)!]^{2}}
		\left(\frac{\beta}{4 S_{1}^{\half}}\right)^{n},
\end{equation}
the last line of which is obtained in Appendix C.  

Thus, for $\alpha = 0$, but $\beta \neq 0$, the free energy is
\begin{eqnarray}
F(\nu;S_{1},S_{2}) & \sim & F_{0}(\nu) + \left[\beta + \frac{1}{2}
		\left(1 - \frac{\beta}{2}\right)\right]\ln S_{1} -
		2\sqrt{S_{1}} + \frac{S_{2}}{2\beta} + \nonumber \\
	&&\left(1 - \frac{\beta}{2}\right)\sum_{n = 1}^{\infty}
		\frac{(2n)!}{[(n + 1)!]^{2}}
		\left(\frac{\beta}{4 S_{1}^{\half}}\right)^{n}.
\label{eq:F-a=0-b-n=2}
\end{eqnarray}

\subsection{The Distribution for $\alpha \neq 0$}

With the effective potential given by Eq.(\ref{eq:u_eff}) and $\alpha
\neq 0$, the charge density is given by
\begin{equation}
\sigma(x) = \frac{1}{\beta\pi}\sfrac{b-x}{x}\left[
	1 - \frac{\alpha}{x + R + \frac{\e}{2}}
		\sfrac{R + \frac{\e}{2}}{b + R + \frac{\e}{2}} -
	\frac{\beta}{x + \e}\sfrac{\e}{b + \e}
	\right],
\label{eq:anz_sigma}
\end{equation}
from which it follows that the normalization condition now reads
\begin{equation}
N - 1 = \frac{b}{2\beta} + \sfrac{\e}{b + \e} -
	\frac{\alpha}{\beta}\left(
		1 - \sfrac{R + \frac{\e}{2}}{b + R + \frac{\e}{2}}
	\right),
\label{eq:anz_norm}
\end{equation}
where, for ease of notation we have introduced $r := R +
\frac{\e}{2}$.  In contrast to the above calculations, we cannot
re-express Eq.(\ref{eq:anz_norm}) in terms of a cubic equation, but
only in terms of an $8th$-order equation, which is why we previously
described this problem as being more difficult than the others.
However, if we define $\nu := 2\beta(N - 1 + \frac{\alpha}{\beta})$,
then we can rewrite Eq.(\ref{eq:anz_norm}) as
\begin{equation}
b = \nu - 2\alpha\sfrac{r}{b + r} - 2\beta\sfrac{\e}{b + \e},
\end{equation}
which we can solve for $b$ as a power series in $\nu$, the result of
the calculation being
\begin{equation}
b = \nu - 2\frac{\alpha r^{\half} + \beta \e^{\half}}{\nu^{\half}} +
	\frac{\alpha^{\frac{3}{2}} + \beta\e^{\frac{3}{2}} - 
		2(\alpha + \beta)(\alpha r^{\half} + \beta\e^{\half})}
	{\nu^\frac{3}{2}} + O\left(\nu^{-\frac{5}{2}}\right),
\end{equation}
which for future reference we write in the compact form $b = \nu +
\delta b$.  

The chemical potential and interaction energy are calculated in
precisely the same manner as before, so we evaluate the former at $x =
b$ and split up the latter into the sum of $F_{1}$ and $F_{\ln}$.
Introducing the further definitions 
\begin{equation}
{\cal R} := \sfrac{r}{b + r} \;\;\;{\rm and}\;\;\;
	{\cal E} := \sfrac{\e}{b + \e}
\end{equation}
then the chemical potential is 
\begin{eqnarray}
A & = & \half\ln(4e)b + r - \alpha\ln(b + r) - \beta\ln[b(b + \e)] + 
	\ln b \left(\alpha(1 - {\cal R}) + \beta(1 - {\cal E}) - 
	\frac{b}{2}\right) + \nonumber \\
	&&	\alpha{\cal R}I_{1}\left(\frac{r}{b}\right) + 
	\beta{\cal E}I_{1}\left(\frac{\e}{b}\right) \\
	& = & \half\ln(4e)b + r - \left(\beta + \frac{b}{2} + 
	\alpha{\cal R} + \beta{\cal E}\right)\ln b + 
	\ln(4) [\alpha {\cal R} + \beta{\cal E}] - \nonumber \\
	&&	2(\alpha\atanh{\cal R} + \beta\atanh{\cal E}).
\end{eqnarray}
Asymptotically we therefore have 
\begin{equation}
F_{\rm chem} \sim F^{0}_{\rm chem}(\nu) - \frac{\sqrt{S_{1}}}{2} -
	\frac{\alpha\sqrt{S_{2}}}{2\beta} + \frac{S_{2}}{4\beta}.
\label{eq:anz_Fchem}
\end{equation}

As for the interaction energy, it is simple to arrive at
\begin{equation}
F_{1} = \frac{b^{2}}{16\beta} - \frac{b}{4\beta}(\alpha{\cal R} +
	\beta{\cal E}) - 
	\frac{\alpha r (1 - {\cal R}) + \beta\e(1 - {\cal E})}
		{2\beta} + 
	\frac{N-2}{2}\big(r - (\alpha + 2\beta)\ln b\big),
\end{equation}
and thus
\begin{equation}
F_{1} \sim F_{1}^{0}(\nu) - \frac{\sqrt{S_{1}}}{2} - 
	\frac{\alpha\sqrt{S_{2}}}{2\beta} + \frac{S_{2}}{4\beta}.
\label{eq:anz_F1}
\end{equation}
In full, the expression for $F_{\ln}$ is
\begin{eqnarray}
F_{\ln} & = & -\left(1 + \frac{\alpha}{\beta}\right)\ln(b)\left(
	\frac{b}{2} - \alpha(1 - {\cal R}) - \beta(1 - {\cal E})
	\right) + \nonumber \\
&&	\frac{b}{2\beta}\left(
	\frac{\beta}{2}\ln(4e) - \alpha I_{3}\left(\frac{r}{b}\right)
	- \beta I_{3}\left(\frac{\e}{b}\right)
	\right) +
	\frac{\beta}{2}\left[{\cal R}I_{2}\left(\frac{\e}{b}\right) + 
		{\cal E}I_{4}\left(\frac{r}{b}\right)\right] +
\nonumber \\
&&	\frac{\alpha}{2}\left[
	{\cal R}I_{4}\left(\frac{r}{b}\right) + 
	{\cal R}I_{5}\left(\frac{\e}{b},\frac{r}{b}\right) +
	{\cal E}I_{5}\left(\frac{r}{b},\frac{\e}{b}\right)\right] +
	\frac{\alpha^{2}}{2\beta}I_{2}\left(\frac{r}{b}\right),
\end{eqnarray}
which yields the asymptotic expansion
\begin{equation}
F_{\ln} \sim F_{\ln}^{0}(\nu) -\sqrt{S_{1}} + \beta\ln(S_{1}) -
	\frac{\alpha}{\beta}\sqrt{S_{2}} + 
	\frac{\alpha}{2}\left(
		1 + \frac{\alpha}{\beta}
	\right)\ln(S_{2}) + 2\alpha \ln(\sqrt{S_{1}} + \sqrt{S_{2}}).
\label{eq:anz_Fln}
\end{equation}

So, from Eqs.(\ref{eq:anz_Fchem}), (\ref{eq:anz_F1}) and
(\ref{eq:anz_Fln}), the asymptotic expansion of the free energy for
non-zero $\alpha$ is
\begin{eqnarray}
F(\nu;S_{1},S_{2}) & \sim & F^{0}(\nu) + \beta\ln(S_{1}) 
	- 
	2\left(\sqrt{S_{1}} + \frac{\alpha}{\beta}\sqrt{S_{2}}\right)
	+ \frac{S_{2}}{2\beta}
	+ \nonumber \\
	&&\frac{\alpha}{2}
		\left(1 + \frac{\alpha}{\beta}\right)\ln(S_{2}) 
	+ 2\alpha\ln(\sqrt{S_{1}} + \sqrt{S_{2}}).
\label{eq:F_a-n=0_b=2}
\end{eqnarray}

\subsection{The Distribution of the Two-Smallest Eigenvalues}
To finally obtain the asymptotic result for the
two-smallest-eigenvalue distribution we first introduce $s_{i} := \nu
x_{i}$, {\it i.e.}, the double-scaled variables of the two smallest
eigenvalues.  Next, we define $P_{\beta}^{(2)}(s_{1},s_{2}) =
\lim_{\nu\rightarrow\infty} p_{\beta}^{(2)}(s_{1}/\nu,s_{2}/\nu)$,
which is the double-scaled asymptotic of the distribution.  From
Eq.(\ref{eq:barp^2}) and the observation that $S_{1} = s_{2} - s_{1}$
and $S_{2} = s_{2}$, we therefore have
\begin{equation}
P_{\beta}^{(2)}(\alpha;s_{1},s_{2}) = {\cal C}
	e^{\beta\ln(s_{2}-s_{1}) - F(\nu;s_{2}-s_{1},s_{2})}.
\label{eq:P^2}
\end{equation}

Explicitly, the asymptotic distribution with $\alpha = 0$, but $\beta
\neq 2$, is
\begin{eqnarray}
P_{\beta}^{(2)}(0;s_{1},s_{2}) & = & {\cal C_{\nu}}
	\exp\left[
	-\frac{1}{2}\left(1 - \frac{\beta}{2}\right)\ln(s_{2}-s_{1}) +
	2\sqrt{s_{2} - s_{1}} - \frac{s_{2}}{2\beta} - \right.\nonumber \\
	&&\left.\left(1 - \frac{\beta}{2}\right)\sum_{n=1}^{\infty}
		\frac{(2n)!}{[(n+1)!]^{2}}\left(
			\frac{\beta}{4(s_{2}-s_{1})^{\frac{1}{2}}}
			\right)^{n}
	\right],
\label{eq:P^2_a=0_b-n=2}
\end{eqnarray}
where ${\cal C_{\nu}}$ represents all the $s_{i}$-independent terms.

Similarly, with $\alpha \neq 0$ and $\beta = 2$, from
Eqs.(\ref{eq:F_a-n=0_b=2}) and (\ref{eq:P^2}), we then have
\begin{eqnarray}
P_{2}^{(2)}(\alpha;s_{1},s_{2}) & = & {\cal C_{\nu}}
	\exp\left[-\frac{s_{2}}{4} + 2\sqrt{s_{2}-s_{1}} +
	\alpha\left(\sqrt{s_{2}} -
	\frac{1}{2}\left(1 + \frac{\alpha}{2}\right)\ln(s_{2})
	\right) - 
	\right. \nonumber	\\
	&&\left.\frac{}{}
	2\alpha\ln\left(\sqrt{s_{2}-s_{1}}\frac{}{}+ 
	\sqrt{s_{2}}\right)\right]
\label{q:P^2_a-n=0_b=2}
\end{eqnarray}

\subsection{A Comparison of the Continuum Approach With Exact
Results}
Exact results for the distribution of the two smallest eigenvalues
have been obtained by Forrester and Huges (FH) \cite{F_and_H} for the
unitary Laguerre ensemble with $\alpha = \ell$, where $\ell$ is either
zero or a positive integer.  We are thus able to check our results and
will do so for the cases $\alpha = 0$ and $\alpha = 1$.  In order to
proceed we first note that in terms of the notation in \cite{F_and_H},
we have $a = 1$ and $m = N$.  Secondly, the variables $s_{i}$
introduced in the previous sub-section are, in the limit of large $N$,
the same as the $s_{i}$ used in \cite{F_and_H}; we are now ready to
check our results.  FH have shown that an exact expression for the
scaled asymptotic distribution is
\begin{eqnarray}
\lefteqn{P^{(2)}_{2}(\ell;s_{1},s_{2}) =  2^{-4}
	e^{-\frac{s_{2}}{4} + \ell \ln(s_{2}/s_{1})}\times }&&\nonumber\\
	&&{\rm det}\left[ \begin{array}{c}
		[I_{j-k+2}(s_{2}^{\frac{1}{2}})]_{\begin{array}{c}
						j=1,\dots,\ell \\
						k=1,\dots,\ell+2\end{array}}
		\\
	\left[\left(\frac{s_{2}-s_{1}}{s_{2}}\right)^{\frac{k-j}{2}}
		I_{j-k+2}(\sqrt{s_{2}-s_{1}})\right]_{\begin{array}{c}
						j=1,2 \\
						k=1,\dots,\ell+2\end{array}}
	\end{array}
	\right],
\end{eqnarray}
which requires us to compute the determinant of an
$(\ell+2)\times(\ell+2)$ matrix of modified Bessel functions of the
first kind, $I_{\nu}(x)$. Note that these are not to be confused
with those in the Table.  

For $\alpha = 0$ and $\alpha = 1$, we thus find that
\begin{equation}
P^{(2)}(0;s_{1},s_{2}) =  \frac{e^{-\frac{s_{2}}{4}}}{16}
	\left[I_{2}\left(\sqrt{s_{2}-s_{1}}\right)^{2} - 
	I_{1}\left(\sqrt{s_{2}-s_{1}}\right)
	I_{3}\left(\sqrt{s_{2}-s_{1}}\right)\right] 
\end{equation}
and
\begin{eqnarray}
\lefteqn{P^{(2)}(1;s_{1},s_{2}) = 
	\frac{e^{-\frac{s_{2}}{4}+\ln(s_{2}/s_{1})}}{16}\times } & &
\nonumber \\
	&&\left\{\left(\frac{s_{2}-s_{1}}{s_{2}}\right)
		I_{2}(s_{2})
	\left[I_{1}\left(\sqrt{s_{2}-s_{1}}\right)^{2} - 
		I_{0}\left(\sqrt{s_{2}-s_{1}}\right)
		I_{2}\left(\sqrt{s_{2}-s_{1}}\right)\right]\right. -
\nonumber \\
	& &\left.\sfrac{s_{2}-s_{1}}{s_{2}}I_{1}(\sqrt{s_{2}})
	\left[I_{1}\left(\sqrt{s_{2}-s_{1}}\right)
	I_{2}\left(\sqrt{s_{2}-s_{1}}\right) - 
		I_{0}\left(\sqrt{s_{2}-s_{1}}\right)
		I_{3}\left(\sqrt{s_{2}-s_{1}}\right)\right] + \right.
\nonumber \\
	& &\left.I_{0}(\sqrt{s_{2}})\left[
		I_{2}\left(\sqrt{s_{2}-s_{1}}\right)^{2} -
		I_{1}\left(\sqrt{s_{2}-s_{1}}\right)
		I_{3}\left(\sqrt{s_{2}-s_{1}}\right)\right]\right\}
\end{eqnarray}
respectively.  We claim that the Coulomb-fluid gives the asymptotic
limit of the given distribution, {\it i.e.}, the distribution in the
limit of $\sqrt{s_{2}-s_{1}},\sqrt{s_{2}} \gg 1$.  In the limit of
large $x$, the modified Bessel functions have the asymptotic expansion
\begin{equation}
I_{\nu}(x) \sim \frac{e^{-x}}{\sqrt{2\pi x}}
\left(1+\frac{4\nu^2-1}{8x}+{\rm O}(x^{-2})\right) 
\end{equation}
The asymptotic
expansion of the given distribution is therefore
\begin{equation}
P^{(2)}_{2}(0;s_{1},s_{2}) \sim 
		\frac{e^{-s_{2}/4 + 2\sqrt{s_{2}-s_{1}}}}{32\pi}
		\left[\frac{1}{s_{2}-s_{1}} +
		O\left((s_{2}-s_{1})^{-2}\right)\right]
\label{eq:P0-ale}
\end{equation}
for $\alpha = 0$, whilst for $\alpha = 1$ we have
\begin{equation}
P^{(2)}_{2}(1;s_{1},s_{2}) \sim  \frac{e^{-s_{2}/4}}{16(2\pi)^{\frac{3}{2}}}
		e^{2\sqrt{s_{2}-s_{1}} + \sqrt{s_{2}}}
	s_{2}^{-\frac{5}{4}}\left[1 + \dots\right].
\label{eq:P1-ale}
\end{equation}
>From the Coulomb-fluid approximation we get
\begin{equation}
P^{(2)}_{2}(0;s_{1},s_{2}) \sim e^{-\frac{s_{2}}{4} +
	2\sqrt{s_{2}-s_{1}}}
\label{eq:P0-ala}
\end{equation}
and 
\begin{equation}
P^{(2)}_{2}(1;s_{1},s_{2}) \sim e^{-\frac{s_{2}}{4} +
	2\sqrt{s_{2}-s_{1}} + \sqrt{s_{2}}}\times 
	e^{- \frac{3}{4}\ln(s_{2}) - 
	2\ln\left(\sqrt{s_{2}-s_{1}} + \sqrt{s_{2}}\right)}.
\label{eq:P1-ala}
\end{equation}
Comparison of Eqs.(\ref{eq:P0-ale}) and (\ref{eq:P0-ala}) and
Eqs.(\ref{eq:P1-ale}) and (\ref{eq:P1-ala}) reveals that the continuum
approximation successfully captures the dominant term in the
asymptotic expansion of the Bessel functions, {\it i.e.}, the term
$e^{-x}$, although the logarithmic factors are not precisely
reproduced.  

\section{Conclusions}

In this paper we have used the continuum approximation to calculate
the GFP, and hence the smallest-eigenvalue distribution, and the
two-smallest-eigenvalue distribution of the Laguerre ensemble of
random matrices.  

The GFP has been the subject of a previous paper \cite{Chen1}, but
here we presented a calculation in which key intermediate results were
given in closed form, in contrast to the simple power series which we
had formerly obtained.

The two-smallest-eigenvalue distribution was new and demonstrates how
it is possible to extend the continuum approximation to deal with
higher-order correlation functions.  The two essential differences in
the calculation being: (i) the normalisation condition is that there
should be $N-2$ charges in the interval and (ii) the set of the
smallest eigenvalues modify the potential in which the remaining
charges are placed, thereby leading to an effective potential instead
of the plain Laguerre potential.  Exact results for this distribution
are available for $\beta = 2$ and $\alpha = 0$ or a positive integer
and we have shown that the continuum approximation gives the leading
order terms in the asymptotic expansion of this distribution for
$\alpha = 0$ and $1$.  We therefore feel justified in claiming that
our result is not only correctly gives the dominant terms for $\alpha
= 2, 3, \dots$, but that the leading-order factor in the asymptotic
expansion of this distribution {\it for all} $\alpha$ and $\beta$ is
given by
\begin{equation}
P^{(2)}_{\beta}(\alpha;s_{1},s_{2}) \sim \exp\left[
	-\frac{s_{2}}{2\beta} + 2\sqrt{s_{2}-s_{1}} + 
	\alpha\sqrt{s_{2}}\right].
\end{equation}
Thus, although the Coulomb-fluid is limited to the dominant factor,
the method is somewhat simpler than the known method of deriving the
distribution exactly \cite{F_and_H} and also has the advantage of being
applicable, we believe, to the full range of parameters of interest.

Since we have shown that the Coulomb-fluid approximation can be
applied rather successfully to problems regarding the smallest and the
two-smallest eigenvalues, it is only natural to ask whether we can use
it examine the $m$-smallest-eigenvalue distribution.  If $x_{1} \leq
x_{2} \leq \dots \leq x_{m}$ denote the $m$ smallest eigenvalues,
then, from Eq.(\ref{eq:eval_dbn}), the $m$-smallest-eigenvalue
distribution can be written in the form
\begin{equation}
p^{(m)}_{\beta}(x_{1},\dots,x_{m}) = {\cal C}^{(m)}_{\beta}
	\prod_{1\leq a < b \leq m}\left|x_{a}-x_{b}\right|^{\beta}
	e^{-\sum_{c=1}^{m}u(x_{c})} {\cal D}^{(m)}_{\beta},
\label{eq:p_m-def}
\end{equation}
where ${\cal C}^{(m)}_{\beta}$ is the normalization constant and 
\begin{eqnarray}
\lefteqn{
	{\cal D}^{(m)}_{\beta} := 
	\int_{x_{m}}^{\infty}dx_{m+1}\dots\int_{x_{m}}^{\infty}dx_{N}
	\left[
	e^{-\sum_{a=m+1}^{N}u(x_{a})}\times\right.} && \nonumber \\
	&&\left.\prod_{1\leq b \leq m, m+1 \leq c \leq N}
		\left|x_{b}-x_{c}\right|^{\beta}
	\prod_{m+1 \leq d < e \leq N}\left|x_{d}-x_{e}\right|^{\beta}
	\right].
\end{eqnarray}
Introducing the change of variable $t_{a-m} = x_{a}-x_{m}$, we have
thus transformed the problem of calculating $p^{(m)}_{\beta}$ to the   
problem of determining ${\cal D}^{(m)}_{\beta}$, where
\begin{equation}
{\cal D}^{(m)}_{\beta} = 
	\int_{0}^{\infty}dt_{1}\dots\int_{0}^{\infty}dt_{N-m}
		e^{-\sum_{a=1}^{N-m}u_{\eff}(t_{a};x_{1},\dots,x_{m})}
	\prod_{1\leq b < c \leq N-m}\left|t_{b}-t_{c}\right|^{\beta},
\end{equation}
the effective potential now being
\begin{equation}
u_{\eff}(t;x_{1},\dots,x_{m}) := u\left(t + x_{m}\right) -
	\beta\sum_{a=1}^{m}\ln\left|t + x_{m} - x_{a}\right|.
\end{equation}
We propose that ${\cal D}_{\beta}^{(m)}$ can be calculated
asymptotically using the Coulomb-fluid approximation with the
normalization being set equal to $N-m$; we will not persue this
problem any further here.
\vspace{10mm}
\par\noindent
{\bf Acknowledgments}

S.M.M. would like to thank the Royal Society for the award of a Royal
Society Reuturn Fellowship which made this research possible.

\newpage
\appendix
\renewcommand{\thesection}{Appendix \Alph{section}.}
\renewcommand{\theequation}{\Alph{section}\arabic{equation}}

\section{The Charging Energy of the Laguerre Ensemble}
\setcounter{equation}{0}

The integral equation for the density, Eq. (9), for $\beta=2$ and
with the specialization to the 
Laguerre Ensemble $(u(x)=x-\ap\ln x)$ reads,
\begin{equation}
x-\ap\ln x-\beta\int_{0}^{b}dy\;\sig(y)\ln|x-y|=A,\;\;
0<x<b.
\label{eq:A1}
\end{equation}
Eq.(\ref{eq:A1}) in addition to regular solutions also admits
solutions that describe point charges.  Including these would produce
infinite self-energy thus causing the total free energy to
diverge. Bearing in mind that the GFP is expressed as the difference
of two free energies, see Eq. (4), the self-energy divergences
cancel. With the point charge at the origin superimposed on the
regular solution, Eq. $(\A1)$ has the following solution:
\begin{equation}
\sigma(x) = -\frac{\ap}{\beta}\delta_{+}(x)+\frac{1}{\pi\beta}
{\sqrt {\frac{b-x}{x}}},\;\;0\leq x<b.
\label{eq:A2}
\end{equation}
Observe that the point charge carries weight of order unity while the
normalization condition $\int_{0}^{b}\sig(x)dx=N,$ is almost exhausted
by the regular distribution.  We find from the normalization
condition, $b=2\beta N+\ap$.  A simple calculation shows that the
number of eigenvalues excluded from the interval, $[0,a],\;(a<b)$ is
$${\cal N}[a]=\int_{0}^{a}\sig(x)dx=
-\frac{\ap}{2\beta}+\frac{1}{\pi\beta}\left[{\sqrt {a(b-a)}}
+b\tan^{-1}{\sqrt {\frac{a}{b-a}}}\;\right].\eqno(\A3)$$
In the scaling limit $b\rightarrow \infty$ and $a\rightarrow 0$
such that $ab=s$ is finite, we find 
$${\cal N}[a]\sim -\frac{\ap}{2\beta}+\frac{2}{\pi\beta}{\sqrt {ab}}
+{\rm O}\left(\frac{a^{\frac{3}{2}}}{b^{\frac{1}{2}}}\right),
\eqno(\A4)$$
and
$$E_{\beta}(s)\sim {\rm exp}\left[-a_1 s+a_2{\sqrt s}\;\right],
\eqno(\A5)$$
where $a_1$ and $a_2$ are constants independent of $s.$
Eq. $(\A5)$ is in qualitative agreement with Eq. (20). As noted in the
introduction the charging energy argument does not supply the logarithmic 
correction.   

\newpage
\section{Evaluation of the Integrals}
\setcounter{equation}{0}
\label{app:integrals}

The integrals $I_{1}$, $I_{2}$ and $I_{4}$ are all evaluated by the
same method, whilst $I_{3}$ is evaluated by elementary means and is
given later.  Firstly, we use the identity $h^{\la}\ln
h=\partial_{\la}\;h^{\la},\; \left({\rm where\;}\partial_{\lambda}=
\frac{\partial}{\partial\la},\right)$ which for the integrands of
interest yields an integral proportional to the Gauss hypergeometric
function \cite{Grad}.  More precisely, we have:
\begin{eqnarray}
I_{1}(x) & = & \frac{1}{\pi}\partial_{\lambda}
\left[B\left(\la,\frac{1}{2}\right)\hy\left(1,\frac{1}{2};
\la+\frac{1}{2};-\frac{1}{x}\right)\right]
\biggl|_{\la=\frac{1}{2}}, \\
I_{2}(x) & = & \frac{1}{2}{\sqrt {\frac{x}{x+1}}}\partial_{\la}
\left[x^{\la}\hy\left(-\la,\frac{1}{2};2;-\frac{1}{x}\right)\right]
\biggl|_{\la=-1}
\end{eqnarray}
and
\begin{equation}
I_4(x)=\frac{1}{\pi x}\partial_{\la}
\left[B\left(\la,\frac{1}{2}\right)\hy\left(1,\la;\frac{3}{2}+\la;
-\frac{1}{x}\right)\right]\biggl|_{\la=\frac{1}{2}}.
\end{equation}
Next, we apply the transformation formul\ae \cite{hyp} 
which send $\hy(.,.,.,1/x)$ to $\hy(.,.,.,x)$ to get a 
convergent hypergeometric series. Since this step and those which 
follow do not essentially differ for the three integrals, we only 
display the calculation for $I_2(x).$ The transformation formula used
is Eq.(9.1322) of \cite{Grad}, from which 
$$I_2(x)={\sqrt {\frac{x}{1+x}}}\left[2\ln\left(\frac{2}{e}\right)
-i_1(x)+\frac{i_2(x)}{{\sqrt x}}\right]+\ln(4x),\eqno(\B4)$$
where
\begin{equation}
i_1(x):=\partial_{\la}\hy\left(1-\la,-\la;\frac{3}{2}-\la;-x\right)
\biggl|_{\la=0}
\end{equation}
and
$$
i_2(x):=\partial_{\la}\left[(1+x)^{\la+\frac{1}{2}}\hy\left(\la,\la+1;
\la+\frac{1}{2};-x\right)\right]\biggl|_{\la=0}.\eqno(\B6)$$
To compute $i_1$ and $i_2$ we use the series representation of
$\hy,$ which is differentiated to give
$$
i_1(x)=-\sum_{n=1}^{\infty}B\left(\frac{3}{2},n\right)(-x)^n,
\eqno(\B7)$$
and
$$
i_2(x)={\sqrt {1+x}}\left[\ln(1+x)+\sum_{n=1}^{\infty}
B\left(\frac{1}{2},n\right)(-x)^n\right].\eqno(\B8)$$
Using the integral representation,
$$B(x,y)=\int_{0}^{1}dt\;t^{x-1}(1-t)^{y-1},\;\;\Re\;x,\;\Re\;y>0,$$
of the Beta function, followed by exchanging the order of 
integration and summation, we find with
the aid of $\sum_{n=0}^{\infty}(-xt)^n=(1+xt)^{-1},\;|xt|<1,$ 
$$i_1(x)=x\int_{0}^{1}dt\frac{{\sqrt {1-t}}}{1+xt},\eqno(\B9)$$
and
$$i_2(x)={\sqrt {1+x}}\left[\ln(1+x)
-x\int_{0}^{1}\frac{dt}{{\sqrt {1-t}}(1+xt)}\right].\eqno(\B10)$$
The integrals in Eqs. $(\B9)$ and $(\B10)$ are expressed in terms of
elementary functions,
$$
i_1(x)=-2+2{\sqrt {\frac{1+x}{x}}}\tanh^{-1}{\sqrt {\frac{x}{1+x}}},
\eqno(\B11)$$
and
$$
i_2(x)={\sqrt {1+x}}\left[\ln(1+x)-2{\sqrt {\frac{x}{1+x}}}
\tanh^{-1}{\sqrt {\frac{x}{1+x}}}\right],\eqno(\B12)$$
and thus the result for $I_2(x)$ follows.\\
To determine $I_3(x),$ we first compute $I_3^{\prime}(x).$ 
The $t-$ integration is then elementary giving
$$
I_3^{\prime}(x)= -1+{\sqrt {\frac{1+x}{x}}}.\eqno(\B13)$$
We find by integrating Eq. $(\B13)$ and supplying a constant,
$$I_3(x)=I_3(0)+\int_{0}^{x}{\sqrt {\frac{1+y}{y}}}dy
=-\frac{\ln(4e)}{2}-x+{\sqrt {x(1+x)}}+\ln\left({\sqrt x}
+{\sqrt {1+x}}\right).\eqno(\B14)$$

The function $I_{5}(x,y)$ requires a little more effort to obtain its
closed form.  We begin by writing the factor $\ln(t+x) = \ln(t) +
P\int_{0}^{x}du/(u+t)$ and thereby obtain
\begin{equation}
I_{5}(x,y) = I_{4}(y) + P\int_{0}^{x}\frac{du}{u-y}
		\int_{0}^{1}\frac{dt}{\pi}\sfrac{1-t}{t}
		\left(\frac{1}{t+y} - \frac{1}{t+u}\right).
\end{equation}
The $t$-integration is easily performed and leads to
\begin{equation}
I_{5}(x,y) = I_{4}(y) + P\int_{0}^{x}\frac{du}{u-y}\left(
		\sfrac{1+y}{y} - \sfrac{1+u}{u}\right).
\end{equation}
Given that
\begin{eqnarray}
\lefteqn{P\int_{0}^{x}\frac{du}{u-y}\sfrac{1+u}{u} =
	\ln\left[1 + 2x + 2\frac{x}{1+x}\right] + }&& \nonumber \\
	&&\sfrac{1+y}{y}\left[\ln(x-y) + \ln\left(-x-y-2xy - 
	2\sqrt{x(1+x)}\sqrt{y(1+y)}\right)\right],
\end{eqnarray}
then the result for $I_{5}$ shown in the Table follows easily.

\newpage

\noindent
\section{Entropic Contribution to the Free Energy}
\setcounter{equation}{0}
\label{app:entropic_term}

The entropic term $f_{\rm ent}^{(2)}$ for the smallest
eigenvalue distribution is given by
$$
f_{\rm ent}^{(2)}=\frac{b-a}{\beta}\int_{0}^{1}
\frac{dt}{\pi}{\sqrt {\frac{1-t}{t}}}
\left(1-\frac{\Lam}{t+x}\right)
\ln\left(1-\frac{\Lam}{t+x}\right),\eqno(\C1)$$
where
$\Lam:=\frac{\ap}{b-a}{\sqrt {\frac{a}{b}}}.$ We could 
evaluate this in closed form, but this would require a method
different to that in Appendix B and as we are most 
interested in the scaling-limit form it is more sensible 
to focus our attention on that. To this end we proceed by first
expanding the logarithm in Eq. $(\C1)$ as a Taylor series in 
$\frac{\Lam}{t+x},$ since ${\sqrt {ab}}>\ap$ 
(the positivity condition on the density) implies
that $\frac{\Lam}{t+x}<1$ for $t>0.$ Doing so gives,
$$f_{\rm ent}^{(2)}=\frac{1}{\beta}\left[-g_1+\sum_{n=1}^{\infty}
\frac{g_{n+1}}{n(n+1)}\right],\;\;
g_1=\ap\left({\sqrt {\frac{a}{b}}}-1\right), \eqno(\C2)$$
where
$$
g_n=\ap{\sqrt {\frac{a}{b}}}\Lam^{n-1}\int_{0}^{1}\frac{dt}{\pi}
{\sqrt {\frac{1-t}{t}}}\left(\frac{1}{t+x}\right)^n
=\ap{\sqrt {\frac{a}{b}}}\frac{(-\Lam)^{n-1}}{(n-1)!}
\left(\frac{d}{dx}\right)^{n-1}{\sqrt {\frac{1+x}{x}}},\;\;n\geq 2.
\eqno(\C3)$$
Writing $(d/dx)^m{\sqrt {1+x}}=u_{m}(1+x)^{\frac{1}{2}-m}$
and $(d/dx)^{m} x^{-\frac{1}{2}}=d_{m}x^{-\frac{1}{2}-m},$ then an 
application of the Liebnitz's rule gives
$$
g_n=(-1)^{n-1}\ap^n\left(\frac{a}{b}\right)^{\frac{n}{2}}
\sum_{p=0}^{n-1}\frac{u_pd_{n-1-p}}{p!(n-1-p)!}\;
b^{\frac{1}{2}-p}\;a^{p-n-\frac{1}{2}}.\eqno(\C4)$$
In the scaling limit we see that the only significant term in
 Eq.(C4) is the $p=0$ one. Since $u_0=1$ (trivially) and
$d_m=(-1)^m(2m)!/(2^{2m}m!),$ then 
$$
g_n=\frac{\ap^n(2n-2)!}{2^{2n-2}\left[(n-1)!\right]^2}\;
\frac{1}{(ab)^{\frac{n-1}{2}}}+
{\rm O}\left(\frac{a}{b(ab)^{\frac{n-1}{2}}}\right)\eqno(\C5)$$
and therefore
$$
f_{\rm ent}^{(2)}\sim -\ap+\frac{1}{\beta}\sum_{n=1}^{\infty}
\frac{\ap^{n+1}(2n)!}{4^n\left[(n+1)!\right]^2}\;
\left(1+\frac{1}{n}\right)\;\frac{1}{s^{\frac{n}{2}}}.\eqno(\C6)$$
The analogous contribution in the two-smallest eigenvalue problem
only differs from the above in that the prefactor is $\frac{b}{\beta},$
as opposed to $\frac{b-a}{\beta},$ $x=\ep/b$ and $\Lam=\frac{\beta}{b}
{\sqrt {\frac{b+\ep}{\ep}}}.$ We therefore now have,
$$
f_{\rm ent}^{(2)}\sim -1+\sum_{n=1}^{\infty}
\frac{(2n)!}{4^n\left[(n+1)!\right]^2}\;\left(1+\frac{1}{n}\right)\;
\left(\frac{\beta}{{\sqrt s}}\right)^n.\eqno(\C7)$$

\newpage

\begin{table}
\begin{tabular}{||c|c|c||} \hline
Function		& Integral Representation	
			& Closed Form \\ \hline
& &\\
$I_{1}(x)$ 
& $\int_{0}^{1}\frac{dt}{\pi}\sfrac{1-t}{t}\frac{\ln(1-t)}{t+x}$ 
& $2\ln 2 + \sfrac{1+x}{x}\left[
	\ln(1+x)-2\atanh\sfrac{x}{1+x}\right]$ \\ 
& &\\\hline
& &\\
$I_{2}(x)$
& $\sfrac{x}{1+x}
	\int_{0}^{1}\frac{dt}{\pi}\sfrac{1-t}{t}\frac{\ln(t+x)}{t+x}$
& $2\ln 2 \sfrac{x}{1+x} + \ln[4x(1+x)] - 
	2\left[1+\sfrac{x}{1+x}\right]\atanh\sfrac{x}{1+x}$\\ 
& &\\\hline
& &\\
$I_{3}(x)$
& $\int_{0}^{1}\frac{dt}{\pi}{\sqrt {\frac{1-t}{t}}}\ln(t+x)$
& $-\frac{\ln(4e)}{2} - x + \sqrt{x(1+x)} +
\ln[\sqrt{x}+\sqrt{1+x}]$\\ 
& &\\\hline
& &\\
$I_{4}(x)$
& $\int_{0}^{1}\frac{dt}{\pi}\sfrac{1-t}{t}\frac{\ln t}{t+x} $
& $2\ln 2 + \sfrac{1+x}{x}\left[\ln x - 2\atanh\sfrac{x}{1+x}\right] $
\\
& &\\\hline
& &\\
$I_{5}(x,y)$
& $\int_{0}^{1}\frac{dt}{\pi}\sfrac{1-t}{t}\frac{\ln(t + x)}{(t+y)}$
& $2\ln 2 - 2\sfrac{1+y}{y}\atanh\sfrac{y}{1+y} -
2\atanh\sfrac{x}{1+x} + $\\
	&	&
	$\sfrac{1+y}{y}\ln\left[
	x + y + 2 x y + 2\sqrt{x(1+x)}\sqrt{y(1+y)}
	\right]$
\\
& &\\\hline
\end{tabular}
\caption{Table of Integrals.  Note that all the resuls are for $x \geq 0$.}
\end{table}


\begin{thebibliography}{99}
\bibitem{gap_prob} Letting $p(x_{1},\ldots,x_{n})$ denote the
probability distribution function of the eigenvalues, then
%
$$
E_{\beta}(x) = \int_{x}^{\infty}dx_{1}\ldots
dx_{n}p(x_{1},\ldots,x_{n}), 	\eqno(1)
$$
%
whilst
%
$$
p^{(1)}_{\beta}(x) = n\int_{x}^{\infty}dx_{2}\ldots
dx_{n}p(x,x_{2},\ldots,x_{n}).
\eqno(2)
$$
%
Eq.(\ref{eq:p_rel_E}) is thus demostrated.
\end{thebibliography}
\end{document}